\documentstyle[prl,aps,psfig,multicol]{revtex}
\begin{document}
\draft
\title{Shubnikov-de Haas oscillations near the metal-insulator
transition in a two-dimensional electron system in silicon}
\author{S.~V. Kravchenko, A.~A. Shashkin$^*$, and David A. Bloore}
\address{Physics Department, Northeastern University, Boston,
Massachusetts 02115}
\author{T.~M. Klapwijk}
\address{Department of Applied Physics, Delft University of
Technology, 2628 CJ Delft, The Netherlands}
\date{\today}
\maketitle
\begin{abstract}
We have studied Shubnikov-de Haas oscillations in a two-dimensional electron system in silicon at low electron densities.  Near the metal-insulator transition, only ``spin'' minima of the resistance at Landau-level filling factors $\nu=$~2, 6, 10, and 14 are seen, while the ``cyclotron'' minima at $\nu=$~4, 8, and 12 disappear.  A simple explanation of the observed behavior requires a giant enhancement of the spin splitting near the metal-insulator transition.
\end{abstract}
\begin{multicols}{2}

There has been a lot of recent interest in the electron properties of dilute two-dimensional (2D) systems, due in part to the observation of an unexpected metal-insulator transition in zero magnetic field \cite{abrahams00}; other puzzling phenomena include the enormous response of these systems to external magnetic field, either perpendicular \cite{diorio90} or parallel \cite{dolgopolov92,simonian97}. In particular, the perpendicular field drives a dilute 2D electron system in silicon metal-oxide-semiconductor field-effect transistors (MOSFETs) through a series of transitions between quantum Hall effect (QHE) states at certain integer filling factors of Landau levels, $\nu$, and insulating states between them, behavior which is described in terms of the metal-insulator phase boundary which oscillates as a function of perpendicular magnetic field \cite{shashkin93}. In this regime, the QHE minima of resistivity have only been seen \cite{diorio90} at filling factors $\nu=$~1, 2, 6, and 10, all of which (except for the one at $\nu=1$) correspond to spin splittings between Landau levels \cite{spin}. No minima have been observed at cyclotron gaps ($\nu=$~4 or 8). This behavior is a puzzle, as the ``cyclotron'' minima of the Shubnikov-de Haas oscillations normally dominate in this 2D system at high electron densities.

In the present paper, we report careful studies of how the Shubnikov-de~Haas oscillations evolve as one approaches the metal-insulator transition in a dilute high-mobility 2D system in silicon. We show that the ``cyclotron'' minima 
gradually disappear near the transition, while the ``spin'' minima remain.  Based on these data, we construct a Landau level fan diagram which maps out the positions of the resistance minima as a function of electron density and magnetic field in the ($B,n_s$) plane.  Within the usual framework, cyclotron minima disappear if the upper spin sublevel of each Landau level coincides in energy with the lower spin sublevel of the next Landau level (within the accuracy of the level broadening).  This requires that, as one approaches the transition from the metallic side, the spin splittings grow by a factor of approximately five relative to their single-particle values until they become nearly equal to $\hbar\Omega_c$ and remain so down to the lowest electron densities (here $\Omega_c$ is the cyclotron frequency).

We used ``split-gate'' samples especially designed for measurements at low electron densities and low temperatures, similar to those previously used in Ref.~\cite{kravchenko00}. In this paper we show results obtained on a sample with a peak mobility $2.3\times 10^4$~cm$^2$/Vs at 4.2~K which increased to $4.0\times 10^4$~cm$^2$/Vs at 0.1~K.

A typical curve taken at low temperature for the longitudinal resistance $R_{xx}$ vs perpendicular magnetic field $B$ is shown in Fig.~\ref{B}(a) for a relatively high electron density (well above the $B=0$ metal-insulator transition which in this sample occurs at $n_s=n_c=8\times10^{10}$~cm$^{-2}$).  Shubnikov-de~Haas oscillations are seen starting at a magnetic field of 0.4~Tesla, and their positions correspond to ``cyclotron'' filling factors, some of which are marked by arrows.  The figure shows that the behavior of the sample at a relatively high electron density is quite ordinary and confirms that this is a uniform and high-quality sample.  In contrast, at low electron density (just above the metal-insulator transition), the magnetoresistance looks quite different as shown in Fig.~\ref{B}(b).  A strong increase in the resistance at $B\gtrsim0.8$~Tesla is caused by the suppression of the metallic behavior at $B=0$ (see Ref.~\cite{kravchenko98}); at higher magnetic fields, its growth is overpowered by the QHE at $\nu=2$ which results in a peak at $B\approx1.3$~Tesla and a minimum of the resistance at $B\approx1.8$~Tesla. In agreement with earlier data \cite{diorio90}, the resistance minima are seen only at $\nu=$~2, 6, and 10 (see the inset); there is also a minimum at $\nu=1$ (not shown in the figure) corresponding to the valley splitting.  There are neither dips nor other anomalies at magnetic fields where cyclotron minima are expected ($\nu=$~4, 8, or 12).

Figure~\ref{ff} shows how the resistance minima corresponding to the cyclotron splittings gradually disappear as the electron density is reduced.  At the highest electron densities (the lower curves), deep resistance minima near even \vbox{
\vspace{10mm}
\hbox{
\hspace{-7mm}
\psfig{file=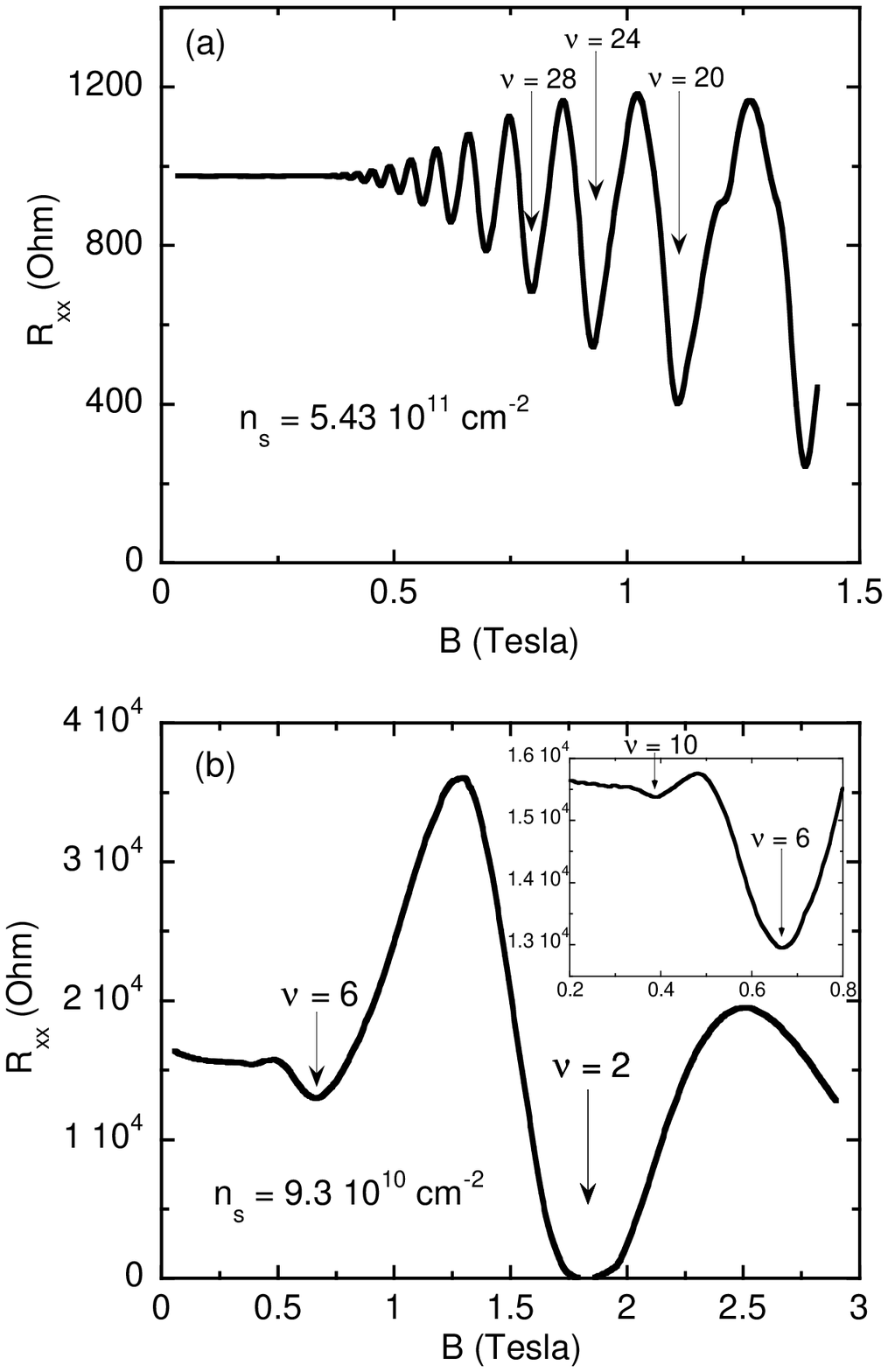,width=3.8in,bbllx=.5in,bblly=1.25in,bburx=7.25in,bbury=9.5in,angle=0}
}
\vspace{-0.55in}
\hbox{
\hspace{-0.15in}
\refstepcounter{figure}
\parbox[b]{3.4in}{\baselineskip=12pt \egtrm FIG.~\thefigure.
Shubnikov-de~Haas oscillations in the Si MOSFET at $T\approx40$~mK (a) at a relatively high electron density $n_s=5.43\times10^{11}$~cm$^{-2}$ and (b) at low electron density $n_s=9.3\times~10^{10}$~cm$^{-2}$.  The minima of the resistance at Landau level filling factors $\nu=6$ and $10$ are shown on an expanded scale in the inset.
\vspace{4mm}
}
\label{B}
}
}
filling factors are seen ($\nu=$~4, 6, and 8 in Fig.~\ref{ff}(a); $\nu=$~10, 12, and 16 in Fig.~2(b)), and a shallow minimum is visible at $\nu=14$ in Fig.~\ref{ff}(b).  As $n_s$ is reduced, the minima at $\nu=$~4, 8, 12, and 16 become less deep, and at the lowest electron densities (the upper curves), neither of them is seen any longer, and only minima at $\nu=$~6, 10, and 14 remain.

The experimental findings are summarized in Fig.~\ref{fan}, where we plot positions of the resistance minima in the ($B,n_s$) plane.  The symbols are the experimental data and the lines are the expected positions of the cyclotron and spin minima calculated according to the formula $n_s=\nu eB/ch$.  The value of the gate voltage, $V_g$, at which the electron density is zero was determined from Shubnikov-de~Haas oscillations in a wide range of electron densities up to $6\times10^{11}$~cm$^{-2}$.  This value is in agreement with that obtained from low-temperature, low-field Hall resistance measurements which will be reported elsewhere \cite{shashkin00}.  One can see that the resistance minima corresponding to spin splittings ($\nu=$~2, 6, 10, 14) extend to much lower electron densities than the ones corresponding to cyclotron splittings ($\nu=$~4, 8, 12).  In this paper,
\vbox{
\vspace{4.25mm}
\hbox{
\hspace{-6mm}
\psfig{file=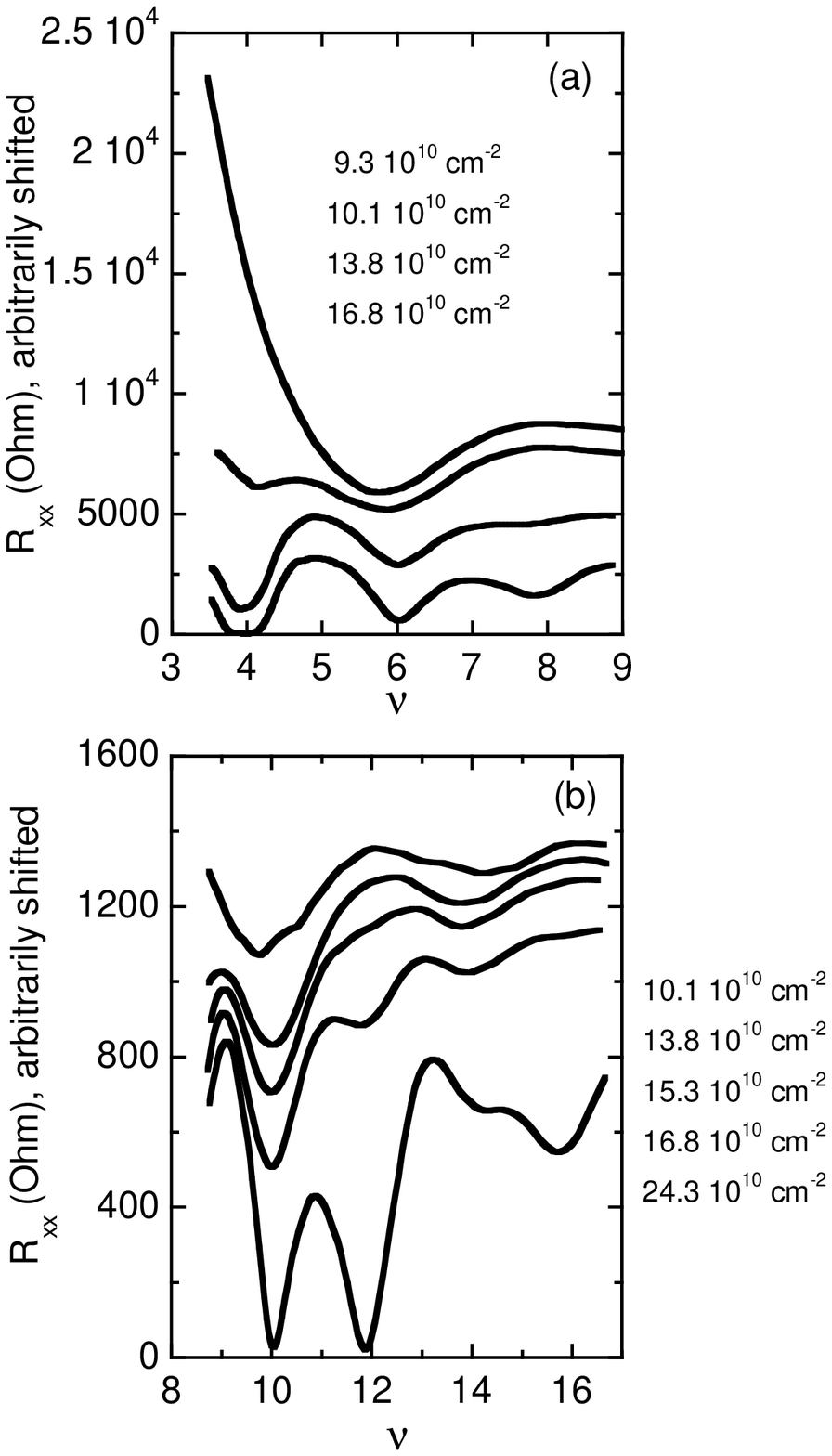,width=3.8in,bbllx=.5in,bblly=1.25in,bburx=7.25in,bbury=9.5in,angle=0}
}
\vspace{-0.15in}
\hbox{
\hspace{-0.15in}
\refstepcounter{figure}
\parbox[b]{3.4in}{\baselineskip=12pt \egtrm FIG.~\thefigure.
Evolution of the Shubnikov-de~Haas oscillations with electron density in two ranges of filling factors: (a) $3<\nu<9$ and (b) $8<\nu<17$; $T\approx40$~mK. The curves are vertically shifted for clarity.
\vspace{0.10in}
}
\label{ff}
}
}
we show only data obtained at electron densities above $n_c$ marked in the figure by the dotted line, although the minima at $\nu=$~2 and 6 do not disappear even when the 2D system becomes strongly localized at $B=0$ \cite{diorio90} ({\it i.e.}, below the dotted line).

At high electron densities, the energy splittings $\Delta_{\text{c}}$ at ``cyclotron'' filling factors ($\nu=$~4, 8, 12...~$=4i$ where $i=$~1, 2, 3...) in Si MOSFETs are much larger than the spin splittings $\Delta_{\text{s}}$ corresponding to filling factors $\nu=$~2, 6, 10...~$=4i-2$.  This is shown schematically in the upper part of Fig.~\ref{fan} (the valley splitting, which is not seen in weak magnetic fields at low electron densities, has been ignored).  Using the effective mass $m=0.19\,m_e$ and the Land\'e $g$-factor $g=2$, one obtains $\Delta_{\text{s}}=g\mu_BB\approx 0.12$~meV/Tesla and $\Delta_{\text{c}}=\hbar\Omega_c-g\mu_BB\approx 0.50$~meV/Tesla (here $m_e$ is the free-electron mass and $\mu_B$ is the Bohr magneton).  Therefore, in the same magnetic field, $\Delta_{\text{s}}$ is expected to be approximately four times smaller than $\Delta_{\text{c}}$, and the Shubnikov-de~Haas minima at $\nu=4i$ should be much deeper than the minima at $\nu=4i-2$.  Our results clearly show that as one approaches the metal-insulator transition, the situation reverses; $\Delta_{\text{c}}$ becomes
\vbox{
\vspace{-9.5mm}
\hbox{
\psfig{file=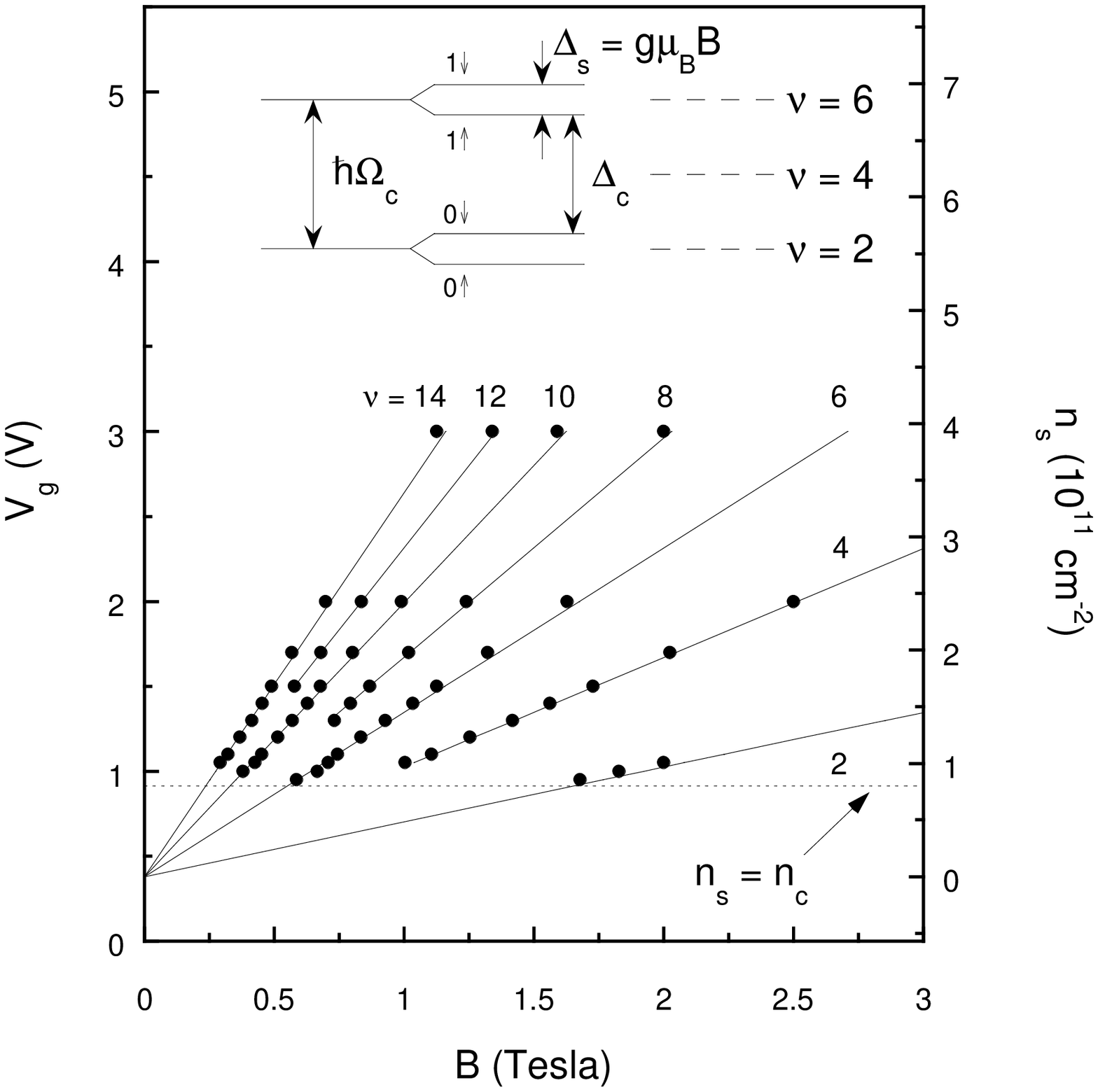,width=2.9in,bbllx=.5in,bblly=1.25in,bburx=7.25in,bbury=9.5in,angle=0}
}
\hbox{
\hspace{-0.15in}
\refstepcounter{figure}
\parbox[b]{3.4in}{\baselineskip=12pt \egtrm FIG.~\thefigure.
Positions of the Shubnikov-de~Haas oscillation minima in the ($B,n_s$) plane (dots) and the expected positions of the cyclotron and spin minima calculated according to the formula $n_s=\nu eB/ch$ (solid lines).  The dotted horizontal line corresponds to the critical electron density for the $B=0$ metal-insulator transition. The upper part of the figure shows the level structure in a Si MOSFET (the valley splittings have been ignored).
\vspace{0.07in}
}
\label{fan}
}
}
smaller than $\Delta_{\text{s}}$ and eventually vanishes.  The condition for vanishing $\Delta_{\text{c}}$ is $g\mu_BB=\hbar\Omega_c$ (within the uncertainty associated with the broadening of the energy levels), or $gm/2m_e=1$, which is higher by more than a factor of five than the ``normal'' value of this ratio, $gm/2m_e=0.19$.

One can attempt to link the observed behavior to a many-body enhancement of spin gaps; in fact, it has been known since early magnetotransport studies of Ref.~\cite{fang68} that the spin gaps exceed the single-particle Zeeman energy. The enhancement of the spin gaps has been explained \cite{ando74} in terms of an additional Coulomb exchange energy $E_{ee}\sim e^2/\epsilon l_B$ that is necessary for creating a spin flip excitation; this has been expressed in terms of an effective exchange-enhanced $g$-factor, $g=g_0+\Delta g$ (here $\epsilon$ is the dielectric constant, $l_B=(\hbar c/eB)^{1/2}$ is the magnetic length, and $g_0=2$ in Si MOSFETs).  The increment $\Delta g$ oscillates in a perpendicular magnetic field between its maximum value, $\Delta g_{\text{max}}$, reached when the Fermi level lies in the middle of the spin gaps ($\nu=4i-2$), and zero when it is outside of the spin gaps. The maximum values of $\Delta g$ depend on magnetic field as
\begin{equation}
\Delta g_{\text{max}}\propto(l_BB)^{-1}\propto B^{-1/2}.\label{eq1}
\end{equation}
At a fixed Landau-level filling factor $\nu=2\pi l_B^2n_s=4i-2$, this
relation can be written as a function of the electron density as
\begin{equation}
\Delta g_{\text{max}}\propto n_s^{-1/2}.\label{eq2}
\end{equation}
Therefore, as $n_s$ (for a fixed filling factor) or $B$ is decreased, $g_{\text{max}}$ should increase \cite{ando74,kallin84}.

The disappearance of the cyclotron splittings in a wide range of magnetic fields, observed in our experiment, requires an enhanced $g$-factor which is independent of magnetic field, in contradiction with Eq.~(\ref{eq1}). Furthermore, according to Ref.~\cite{kallin84}, {\em all} energy gaps including $\Delta_c$, rather than only spin gaps, should be many-body enhanced by approximately the same amount $E_{ee}$ when the Fermi level lies in the middle of the gaps.  This makes the mechanism considered above for the disappearance of the cyclotron gaps even less plausible.

On the other hand, our results seem consistent with the suggestion \cite{okamoto99} that in relatively low magnetic fields, of the order of those used in our experiment, the average $g$-factor is nearly field-independent and approximately equal to its many-body enhanced zero-field value.  In experiments \cite{fang68,okamoto99}, the $g$-factor was measured by a comparative determination of spin gaps using tilted magnetic fields (the so-called beating pattern method which yields some average $g$-factor between $g_0$ and $g_{\text{max}}$). In Ref.~\cite{okamoto99}, an empirical relation between $n_s$ and the ratio $gm/2m_e$ was established and it was shown that the latter is approximately proportional to $n_s^{-1/2}$.  Extrapolating this relation to $n_s=1\times10^{11}$~cm$^{-2}$, one obtains the ratio $gm/2m_e\approx0.6$ which is still not large enough for the cyclotron splittings to disappear.  The appreciably higher ratio $gm/2m_e\approx1$ necessary to explain our results at $n_s\sim10^{11}$~cm$^{-2}$ suggests that the $g$-factor grows more rapidly with decreasing electron density in the vicinity of the metal-insulator transition. The enhancement of the $g$-factor at low electron densities due to electron-electron interactions is expected within Fermi-liquid theory \cite{iwamoto91}, although it is not known whether it can be as large as we report in this paper (a factor of $\sim5$).  It is also unclear why, once the spin splittings reach $\sim\hbar\Omega_c$, they remain so down to the lowest electron densities.

In summary, we have shown that in a low-density 2D electron system in silicon, the Shubnikov-de~Haas oscillation minima corresponding to the cyclotron splittings disappear as one approaches the metal-insulator transition, and only spin minima survive.  This behavior is observed in a rather broad range of magnetic fields.  Within the usual theoretical framework, this requires an unexpectedly strong enhancement of the $g$-factor at electron densities close to $n_c$.

We gratefully acknowledge discussions with M.~P. Sarachik and S.~A. Vitkalov.  This work was supported by NSF grants DMR-9803440 and DMR-9988283 and Sloan Foundation.




\end{multicols}

\begin{references}
\bibitem[*]{}Permanent address: Institute of Solid State Physics,
Chernogolovka, Moscow District 142432, Russia.
\bibitem{abrahams00} E. Abrahams, S.~V. Kravchenko, and M.~P. Sarachik, preprint cond-mat/0006055.
\bibitem{diorio90} M. D'Iorio, V.~M. Pudalov, and S.~G. Semenchinsky, Phys.\ Lett.\ A\ {\bf 150}, 422 (1990); V.~M. Pudalov, M. D'Iorio, and J.~W. Campbell, Surf.\ Sci.\ {\bf 305}, 107 (1994).
\bibitem{dolgopolov92} V.~T. Dolgopolov, G.~V. Kravchenko, A.~A. Shashkin, and S.~V. Kravchenko, JETP Lett.\ {\bf 55}, 733 (1992).
\bibitem{simonian97} D. Simonian, S.~V. Kravchenko, M.~P. Sarachik, and V.~M. Pudalov, Phys.\ Rev.\ Lett.\ {\bf 79}, 2304 (1997); V.~M. Pudalov, G. Brunthaler, A. Prinz, and G. Bauer, JETP Lett.\ {\bf 65}, 932 (1997).
\bibitem{shashkin93} A.~A. Shashkin, G.~V. Kravchenko, and V.~T. Dolgopolov, JETP\ Lett.\ {\bf 58}, 220 (1993).
\bibitem{spin} In silicon MOSFETs, gaps between spin-up and spin-down levels correspond to $\nu=$~2, 6, 10, 14... due to a two-fold valley degeneracy in this system.
\bibitem{kravchenko00} R. Heemskerk and T.~M. Klapwijk, Phys.\ Rev.\ B {\bf 58}, R1754 (1998); S.~V. Kravchenko and T.~M. Klapwijk, Phys.\ Rev.\ Lett.\ {\bf 84}, 2909 (2000).
\bibitem{kravchenko98} S.~V. Kravchenko, D. Simonian, M.~P. Sarachik, A.~D. Kent, and V.~M. Pudalov, Phys.\ Rev.\ B\ {\bf 58}, 3553 (1998).
\bibitem{shashkin00} A.~A. Shashkin {\it et al.}, to be published.
\bibitem{fang68} F.~F. Fang and P.~J. Stiles, Phys.\ Rev.\ {\bf 174}, 823 (1968).
\bibitem{ando74} T. Ando and Y. Uemura, J.\ Phys.\ Soc.\ Jpn.\ {\bf 37}, 1044 (1974).
\bibitem{kallin84} Yu.~A. Bychkov, S.~V. Iordanskii, and G.~M. Eliashberg, JETP Lett.\ {\bf 33}, 143 (1981); C. Kallin and B.~I. Halperin, Phys.\ Rev.\ B\ {\bf 30}, 5655 (1984); A.~H. MacDonald, H.~C.~A. Oji, and K.~L. Liu, Phys.\ Rev.\ B {\bf 34}, 2681 (1986).
\bibitem{okamoto99} T. Okamoto, K. Hosoya, S. Kawaji, and A. Yagi, Phys.\ Rev.\ Lett.\ {\bf 82}, 3875 (1999).
\bibitem{iwamoto91} See, {\it e.g.}, N. Iwamoto, Phys.\ Rev.\ B {\bf 43}, 2174 (1991); Y. Kwon, D.~M. Ceperley, and R.~M. Martin, Phys.\ Rev.\ B {\bf 50},
1684 (1994).

\end{references}
\end{document}